\documentstyle[12pt]{article}
\begin{document}
\setlength{\unitlength}{1mm}
{\hfill Preprint JINR E2-93-33 } \vspace*{2cm} \\
\begin{center}
{\Large\bf Two-dimensional Black Hole With Torsion}
\end{center}
\begin{center}
{\large\bf S.N.Solodukhin}
\end{center}
\begin{center}
{\bf Laboratory of Theoretical Physics,
Joint Institute for Nuclear Research,
Head Post Office, P.O.Box 79, Moscow, Russia}
\end{center}
\vspace*{2cm}
\abstract
   The 2D model of gravity with zweibeins $e^{a}$ and the Lorentz
connection one-form $\omega^{a}_{\ b}$ as independent gravitational variables
is considered and it is shown that the classical equations of motion
are exactly integrated in coordinate system determined by components of
2D torsion. For some choice  of integrating constant the solution is of
the charged black hole type. The conserved charge and ADM mass of  the
black hole are calculated.
\vspace*{2cm}
\newpage
   Recently much attention has been payed to the investigation of
two-dimensional dilaton gravity. This is mainly inspired  by string theory,
and also by the fact that it gives the simplest model for the dynamical
description of a
two-dimensional gravity [1-5], the gravitational variables are dilaton
and metric fields $( \phi , g_{\mu  \nu} ) $. In empty  (without matter)
space the classical equations of motion are exactly integrated [1-3] and
the solution describes the two-dimensional black hole. On the quantum level
it was shown [4] that this model is renormalizable. Since in two dimensions
many things are simpler and models (classical and quantum) become solvable,
one can consider the 2D dilaton gravity as "toy model" for the study of old
problems of black hole formation and evaporation [5].

   On the other hand, numerous recent attempts to formulate the
theory of gravity in the framework of a consistent gauge approach resulted in
constructing the gauge gravity models for the de Sitter and Poincar\'e groups
(for a review
see, e.g., [6] ). The independent variables are now vielbeins
$e^{a}= e^{a}_{\mu}dx^{\mu}$ and Lorentz connection one-form
$\omega^{a}_{\ b} = \omega^{a}_{\ b,\mu} dx^{\mu}$.
The application of these methods in two dimensions was
justified by attempts to give an alternative description of two-dimensional
dynamical gravity in terms of variables $( e^{a} , \omega^{a}_{\ b} )$. It
was argued also that investigation of simple two-dimensional model leads to
a better understanding of four-dimensional gravity and its quantization [7].
It was shown in [7] that the Lagrangian $L= \gamma R^{2} + \beta
T^{2} + \lambda $ is the most general one quadratic in curvature
$R$ and torsion
$T$, and containing a cosmological constant $\lambda$. The classical equations
of motion were analyzed in conformal gauge [7] and in light cone gauge [8] and
their exact integrability was demonstrated.

   In this note  we will consider the model for two dimensional de Sitter
gravity. The constants $\gamma, \beta, \lambda$ are fixed in this case with
only one free parameter $\alpha^{2}$ and the action is of
the Yang-Mills type [6]. We will show that the exact solution of equations of
motion is most easily found in coordinates given by torsion's components (the
generalization of result to the case of arbitrary constants $\gamma, \beta,
\lambda$ is straightforward). For certain choice of integrating constant this
solution is of the black hole type.

\bigskip

   {\bf 1}. In two dimensions the gauge gravity is described in terms of
zweibeins
$e^{a} = e^{a}_{\mu} dz^{\mu}, a=0,1 $ (the 2D metric on the surface $M^{2}$
has the form $g_{\mu \nu}=e^{a}_{\mu} e^{b}_{\nu} \eta_{ab} $) and Lorentz
connection one-form $\omega^{a}_{\ b} = \omega \varepsilon^{a}_{\ b}, \  \omega
=
\omega_{\mu} dz^{\mu} \  (\varepsilon_{ab} =- \varepsilon_{ba}, \
\varepsilon_{01}=1)$.
The de Sitter curvature two-form ${ \cal R}$ [6] in two dimensions takes the
form:
\begin{displaymath}
{\bf {\cal R}}=
\left( \begin{array}{ccc}
R \epsilon^{a}_{\ b} + \alpha^{2} e^{a} \wedge e_{b} & \alpha T^{a} \\
\alpha T_{b} & 0 \\
\end{array} \right)
\end{displaymath}
where $\alpha$ is the coupling constant, and curvature and torsion two-forms
are: $R=d\omega, \ T^{a}=de^{a} + \varepsilon^{a}_{\ b}
\omega \wedge e^{b}$.

   The dynamics of gravitational variables $( e^{a}, \omega)$ is determined by
the action of the Yang-Mills type [6]:
\begin{eqnarray}
&&S=\int\limits_{M^{2}}^{} {1 \over 4} Tr \ast {\cal R} \wedge {\cal R}
\nonumber \\
&&= \int\limits_{M^{2}}^{} {\alpha^{2} \over 2} \ast T^{a} \wedge T^{a} +
{1 \over 2} \ast R \wedge R - {\alpha^{4} \over 4} \varepsilon_{ab} e^{a} e^{b}
+  \alpha^{2}  R
\end{eqnarray}
where $\ast$ is the Hodge dualization. The last term in (1) is the boundary one
 and it does not affect the
equations of motion.

   Let us consider variables $\rho = \ast R$ and $q^{a} = \ast T^{a}$.
Then using identity $\varepsilon_{ab} \varepsilon^{cd} =(\delta^{d}_{a}
\delta^{c}_{b} - \delta^{c}_{a} \delta^{d}_{b})$ it
is easy to show that the action (1) takes the form:
\begin{center}
$S=-{1 \over 4} \int\limits_{}^{} [\alpha^{2} q^{2} + (\rho + \alpha^{2}
)^{2} ] e^{a} \varepsilon_{ab} e^{b},$ \\
\end{center}
which is positive in euclidean signature;
here $q^{2}=q^{a}q^{b} \eta_{ab} \ (\eta_{ab}=diag(+1,-1))$.

   Variation of action (1) with respect to zweibeins $e^{a}$ and Lorentz
connection $\omega$ leads to the following equations of motion:
\begin{equation}
d \rho =- {\alpha}^{2} q^{a}\varepsilon_{ab}e^{b}
\end{equation}
\begin{equation}
\nabla q^{a} = -{1 \over 2{\alpha}^{2} } [\rho^{2} +\alpha^{2} q^{2}
- \alpha^{4}] \varepsilon^{a}_{\ b} e^{b},
\end{equation}
where $\nabla q^{a} \equiv dq^{a} + \omega \varepsilon^{a}_{\ b}q^{b}$.

\bigskip

   {\bf 2}. One particular solution of (2)-(3) is evident.
Assuming $q^{2}=constant$ one gets from (2)-(3), provided
$e^{a}$ are linearly independent everywhere on $M^{2}$:
\begin{center}
$\rho^{2} = \alpha^{4}, \ q^{a}=0 $ \\
\end{center}
in all points of the two-dimensional manifold. That is, torsion is zero
and $M^2$ is the de Sitter space.

   Let now $q^{2}$ be nonconstant, and hence non zero identically everywhere
in $M^{2}$. Then from
eqs.(2)-(3) we have the following equation connecting $q^{2}$ and $\rho$:
\begin{equation}
{dq^{2} \over d\rho} = { 1 \over \alpha^{4}} \Phi,
\end{equation}
where $\Phi (\rho, q^{2}) = \rho^{2} + \alpha^{2} q^{2} - \alpha^{4}$.

The solution of this equation has the form:
\begin{equation}
q^{2}(\rho) = - {1 \over \alpha^{2}} (\rho + 2\alpha^{2})^{2}
+ \epsilon e^{\rho \over \alpha^{2}},
\end{equation}
where $\epsilon $ is integrating constant, we will see that it is proportional
to ADM mass. Notice that due to pseudoeuclidean signature $q^{2}$ can take
both positive and negative values.

   One can see that for large negative $\rho$ independently of the value of
integration constant $\epsilon$ function $q^{2}(\rho)$ has the  asymptotics
$q^{2} \sim  - {1 \over \alpha^{2}} (\rho + \alpha^{2}  )^{2}$.
The form of this function for positive $\rho$ depends on the constant
$\epsilon$.

   {\bf A. $\epsilon > 0$}

   In this case  for large positive $\rho$ function $q^{2}$ is positive and
approximately $q^{2} \sim \epsilon e^{\rho \over \alpha^{2}}$.

   The  critical points of function $q^{2}(\rho)$  (5)
(where ${dq^{2} \over d\rho}=0$) are solutions of following equation:

\begin{equation}
\rho_{c} = - \alpha^{2} + {\epsilon \over 2}e^{\rho_{c} \over \alpha^{2}}
\end{equation}
One can show that there are no such points for $\epsilon > 2\alpha^{2}$;
for $\epsilon = 2\alpha^{2}$ one gets one critical point
$\rho_{c}=0$; for $0 < \epsilon < 2\alpha^{2}$ the function has
two critical points: the first one is positive $(\rho_{c1} > 0)$ and the
second is negative $(\rho_{c2} < 0)$.

   In general case $q^{2}(\rho)$ in critical point is equal to the following
value:
\begin{equation}
q^{2}_{c} = {\epsilon \over 2} e^{\frac{\rho_{c}}{\alpha^{2}}}
(1-{\rho_{c} \over \alpha^{2}})
\end{equation}
One can see that  $q^{2}_{c}$ is positive if $\rho_{c} <0$ (since $\epsilon
>0$). The sign of $q^{2}$ in positive critical point $\rho_{c1}$ depends on
value of constant $\epsilon$. If $\epsilon$ is slightly smaller than
$2\alpha^{2}$  then $q^{2}_{c1}$ is still positive. The point
$\rho_{c1}$ is a minimum  which  goes down with decreasing constant
$\epsilon$ and reaches zero value $q^{2}_{c1}=0$ if, as follows
from (7), $\rho_{c1}=\alpha^{2}$. One can see from (6)
that it corresponds to $\epsilon={4\alpha^{2} \over e}$ \footnotemark\
\addtocounter{footnote}{0}\footnotetext{$e=2.7...$
 is the Euler number}. Thus we come to
following conclusion about the behavior of function $q^{2}(\rho)$.

For $\epsilon > {4\alpha^{2} \over e}$ the function $q^{2}(\rho)$ has only
one zero at a negative $\rho <- \alpha^{2}$.
If $\epsilon={4\alpha^{2} \over e}$ there are two such zeros:
at $\rho<- \alpha^{2}$ and $\rho=\alpha^{2}>0$.
For $0<\epsilon < {4\alpha^{2} \over e}$ the function $q^{2}(\rho)$ vanishes
at three points: one for $\rho <- \alpha^{2}$ and two
for $\rho >-\alpha^{2}$ (one of which satisfies $\rho > \alpha^{2}$).

   {\bf B. $\epsilon=0 $}

   In this case function (5) reduces to $q^{2}= -\frac{1}{\alpha^{2}}
(\rho + \alpha^{2})^{2}$ which is negative everywhere except for a point
$\rho=-\alpha^{2}$ where it vanishes.

   {\bf C. $\epsilon < 0 $}

   As one can see from (5) the function $q^{2}(\rho)$ has no zeros in this
case and it is negative for any $\rho$. Evidently there is only one
critical point (maximum) $\rho_{c}$ which lies in the interval
$-\alpha^{2} - \frac{|\epsilon|}{e} < \rho_{c} < - \alpha^{2}$.

\bigskip

   {\bf 3.} Thus eqs.(2)-(3) determine $q\ (=\sqrt{q^2})$ as a function of
$\rho$. Further analysis of (2) easily shows that $\xi (q)=0$, where we
denoted 1-form $\xi = q_{c} e^{c}$. Let us use this and introduce a new
coordinate system which is (pseudo)polar with $q$ playing the role of a
'radial' coordinate, while the 'angular' coordinate $\phi$ is then clearly
such that its differential is proportional to $\xi$. Assuming (for
definiteness) that $q^{2}=(q^{0})^{2}- (q^{1})^{2} >0$, one can write the
torsion components in the form: $q^{0}=q \cosh \phi, \ q^{1}=q \sinh \phi$.

   Let us consider $q,\phi$ as the new local coordinates on $M^2$. The
differentials $\{ dq, d\phi \}$ form basis in the space of one-forms.
Since $q$ is a function of $\rho$, we can use an equivalent basis
$\{d\rho, \ d\phi \}$. From the construction of $q,\phi$ (see above)
and (2)-(3) we get
\begin{eqnarray}
&&q^{a}\varepsilon_{ab} e^{b}= - \frac{d\rho}{\alpha^{2}} \nonumber \\
&&q_{a} e^{a} = \xi = Bd\phi
\end{eqnarray}
where $B$ is some function of variables $\rho$ and $\phi$. Equations (8) are
easily solved with respect to zweibeins $e^{a}$:
\begin{eqnarray}
&&e^{0}= {B \over q}\cosh{\phi}d\phi -\frac{1}{\alpha^{2}q} \sinh {\phi} d\rho
\nonumber \\
&&e^{1}= {B \over q}\sinh{\phi}d\phi -\frac{1}{\alpha^{2}q} \cosh {\phi} d\rho
\end{eqnarray}
Let us find the function $B$. From (9) one calculates the volume 2-form
$V={1 \over 2}\varepsilon_{ab}e^{a} \wedge e^{b}={B \over q^2 \alpha^2}d\rho
\wedge
d\phi$, and from (8) thus $d\xi = q^2 \alpha^2 B^{-1}{\partial B \over
\partial \rho} V$. However, it is straightforward to see that (3) can be
rewritten as
$$
d \xi = ({\Phi \over \alpha^2} - q^2)V =
$$
$$
=q^2 \alpha^2 (q^{-2}{dq^2 \over d\rho} - {1 \over \alpha^2})V,
$$
and thus $B$ satisfies
$$
B^{-1}{\partial B \over \partial \rho}=q^{-2}{dq^2 \over d\rho} -
{1 \over \alpha^2}.
$$
{}From this we find finally
\begin{center}
$B=q^{2}B_{0} \exp{(-\frac{\rho}{\alpha^{2}})} $ \\
\end{center}
where $B_{0}$ is an arbitrary function of $\phi$. Consequently the metric has
the form:
\begin{equation}
d^{2}s= (e^{0})^2 - (e^{1})^2 =
q^{2}(\rho) \exp{(-\frac{2\rho}{\alpha^{2}})}(d\phi)^{2}
-\frac{1}{\alpha^{4} q^{2}(\rho)}(d\rho)^{2}
\end{equation}
where $q^{2}(\rho)$ is known function (5), and without any loose of generality
we redefined the 'angular' variable $B_{0}(\phi)d\phi \rightarrow d\phi$
(denoting the new coordinate by the same letter). Several remarks are in
order. First of all, let us note that this result (10)
is also valid in region where $q^{2} < 0$ (in which case one should use other
formulas describing the introduction of new coordinates, $q^{0}=q \sinh \phi$,
$q^{1}=q \cosh \phi$). Secondly, the metric (10) is obviously stationary (angle
$\phi$ is of course the time coordinate), however the Lorentz connection
one-form $\omega$ has a non-stationary part due to the time-dependent torsion.
It should be noted also that the demonstration of integrability of the model
(1) in coordinates of torsion $q^{a}$ is similar to the analysis of the dilaton
gravity without matter in the coordinate system where the dilaton field
$\phi$ plays the role of one of coordinates [1]. As was shown in [7], in
conformal gauge the action of the type (1) leads to the essentially nonlinear
equations of motion, the exact solution of which is much more complicated.
Thus the conformal gauge, though standard, is not always the best one.
The model (1) gives us an example of a theory when the equations dictate the
natural gauge.

\bigskip

    {\bf 4}. It is straightforward to see that the above
results are valid (with slight appropriate modifications) also for the theory
with the most general action, $S=\int L V$,
\begin{equation}
L = \beta T^{a}_{\ \mu\nu}T_{a}^{\ \mu\nu} + \gamma R_{\mu\nu}
R^{\mu\nu} + \lambda,
\end{equation}
where $R_{\mu \nu}= \partial_{\mu} \omega_{\nu} - \partial_{\nu} \omega_{\mu}$.

Equations (2), (3) are then replaced by the generalized equations of motion
\begin{equation}
d \rho =- {\alpha}^{2} q^{a}\varepsilon_{ab}e^{b}
\end{equation}
\begin{equation}
\nabla q^{a} = -{1 \over 2{\alpha}^{2} } [\rho^{2} +\alpha^{2} q^{2}
- \alpha^{4} - \Lambda^2] \varepsilon^{a}_{\ b} e^{b},
\end{equation}
where the following notation was introduced:
$$
\alpha^2={\beta \over \gamma}, \ \ \Lambda^2= - {{\lambda \over 2\gamma} -
{\beta^2
\over \gamma^{2}}}.
$$
Notice that squares were introduced in order to get formal similarities
with the above considered de Sitter model, but actually neither newly defined
$\alpha^2$ nor $\Lambda^2$ are positive, in general. As is seen from (1), the
de Sitter model is recovered when $ \lambda = -
2\beta^{2} / \gamma$, so that $\Lambda=0$ and old and new $\alpha$'s coincide.

Equations (12), (13) yield a modified equation for $q^2$ which replaces (4),
\begin{equation}
{dq^{2} \over d\rho} = { 1 \over \alpha^{4}} \Phi^{\prime},
\end{equation}
where $\Phi^{\prime}(\rho, q^{2}) = \rho^{2} + \alpha^{2} q^{2} - \alpha^{4}
- \Lambda^2$.

The solution of this equation has the form:
\begin{equation}
q^{2}(\rho) = - {1 \over \alpha^{2}} (\rho + \alpha^{2})^{2}
+ {\Lambda^2 \over \alpha^2}
+ \epsilon e^{\rho \over \alpha^{2}}.
\end{equation}
The behavior of this function is very similar to (5); in fact if one
confines to the case $\alpha^2 > 0$, the quantitative analysis of $q^2$
almost duplicates the discussion in sect.2.

It is easy to see that equations (12)-(15) are such that the introduction
of the $q,\phi$ coordinates proceeds precisely as described in the sect.3,
including the same solution for the function $B$.

Hence, the general theory (11) has the same structure of the line-element
(10), with $q^2$ determined from (15).

\bigskip

    {\bf 5.} The coupling with matter in general case breaks this exact
integrability in coordinates $q^{a}$. The exception is the coupling with
Yang-Mills fields. In this case the total action
$$
S=\int\limits_{M^{2}}^{} {1 \over 4} (Tr \ast {\cal R} \wedge {\cal R}
+ Tr \ast F \wedge F),
$$
where $F=dA + A \wedge A $ is strength of gauge field $A=A^{a}_{\mu}
\tau^{a}dz^{\mu}$, yeilds the equations of motion for the gauge field
 $A^{a}_{\mu}$ which reduce to the conservation law $df^{2}=0$,
$f^{2}= Tr[\ast F\ast F]$. This means that $f^{2}=constant $; in abelian case,
moreover, $f= \ast F = constant$ is a charge. In equations (2)-(3) the coupling
of the
gauge gravity with the Yang-Mills matter is manifested only in the shift of
"cosmological" constant $\lambda=-2\alpha^{4}$ as $\lambda \rightarrow
-2\alpha^{4} + f^{2}$. Correspondingly the function $q^{2}$ changes:
$q^{2} \rightarrow q^{2} - \frac{f^{2}}{2\alpha^{2}}$ so that the metric again
has the form (10).

\bigskip

   {\bf 6.} Let us remind that when varying the action in order to get an
equation of motion one usually drops out the surface term which arises when
integrating by parts. The correct way of doing so is to impose appropriate
boundary conditions. Assuming the variations $\delta \omega$ and $\delta
e^{a}$ are arbitrary at spatial infinity (which in the polar coordinate
system corresponds to the infinite value of the usual radial coordinate)
one gets from the action (1) the boundary conditions:
\begin{equation}
\rho |_{\infty} =-\alpha^{2}; \ \ q^{a}|_{\infty}=0
\end{equation}

The constraint  that torsion at space infinity is zero is
too strong. It leads to the constraint $\epsilon=0$ in (5), so most of
solutions  are omitted.

Let us add to the action (1) following term:
\begin{equation}
S_{b}= - a \alpha^{2} \int_{\partial M^{2}} \gamma^{1 \over 2} d\tau
\end{equation}
where $\gamma=det \gamma_{\mu \nu}, \ \gamma_{\mu \nu}=g_{\mu \nu} -
\kappa n_{\mu}n_{\nu}$ is metric induced on the boundary $\partial M^{2}$
with normal vector $n_{\mu} \ (n_{\mu} n_{\nu} g^{\mu \nu}= \kappa)$, $\kappa
=1$ for spacelike boundary and $\kappa=-1$ for timelike boundary. Then
variation of total action $S_{tot}=S+S_{b}$ (note that it is still positive
in euclidean signature for $a>0$) leads to the modified boundary conditions at
space
infinity, which for metric (9)-(10) take the form
\begin{equation}
\rho |_{\infty} =-{\alpha^{2} }; \ \ q|_{\infty}=a
\end{equation}

It means that integrating constant in (5) $\epsilon =a^{2}e$. Thus the found
solution describes the two-dimensional asymptotically de Sitter space with
two kinds of possible singularities: where $\rho=- \infty$ and $\rho=+\infty$.
Remember that $\rho$ is curvature of the Lorentz connection one-form
$\omega$: $\rho= \ast(d \omega)$. One can show that relationship of $\rho$
 and "metrical" curvature $\rho_{0}= \ast[d(\ast(de^{a})e^{a})]$
determined for metric (10)  is given by the formula:
\begin{equation}
\rho_{0}= \rho -{1 \over \alpha^{2}} \rho^{2} + \alpha^{2}
\end{equation}
So we have $\rho_{0}=\rho$ if $\rho=\pm {\alpha^{2} }$ and singularity
of $\rho$ means the singularity of $\rho_{0}$, one can see from (19) that at
any kind of singularity $\rho_{0}$ is negative.

\bigskip

   {\bf 7.} The most interesting solution is of the type {\bf A} with
$\rho$ laying in the interval $-{\alpha^{2} } \leq \rho < +\infty$.
One can see that metric (10) describes the two-dimensional asymptotically
de Sitter
space-time with singularity ($\rho=+\infty$) and horizons at points where the
function $q^{2}(\rho)$ has zeros.

As was described above, for $\epsilon >
{4\alpha^{2} \over e}$ such points are absent and we have naked singularity.
For $0 < \epsilon < {4\alpha^{2} \over e}$ we obtain two horizons which
coincides when $\epsilon={4\alpha^{2} \over e}$. Thus the metric (10) for
$0< \epsilon \leq {4\alpha^{2} \over e}$ resembles the charged two-dimensional
black hole type solution [3] (though  the (10) is not exactly metric considered
in [3] for 2D dilaton gravity coupled with Maxwell field) .
The case $\epsilon={4 \alpha^{2} \over e}$
corresponds to the extremal black hole.

   In support of this analogy we note that the equation (2) is similar to
the  Maxwell equation $df= \ast j$
where  $f= \ast (dA)$ is strength of abelian gauge field $A$ and $j$ is charged
matter current one-form. Then the second gravitational equation (3) is
similar to the equation of motion for charged matter. It is not surprising
because the local Lorentz symmetry in two dimensions is abelian and
analogous to the $U(1)$-symmetry of Maxwell theory.

   From eq.(3) we get that the corresponding Lorentz current one-form $\ast
J= - \alpha^{2} q^{a} \varepsilon_{ab} e^{b}$ is conserved, $d \ast J=0$.
Integrating $\ast J$ over any spacelike hypersurface $\Sigma$ we get that total
charge $Q=\int_{\Sigma} \ast J$ is equal to curvature $\rho$ at
infinity\footnotemark\addtocounter{footnote}{0}\footnotetext
{Note that the same formula is valid in (1+1)-electrodynamics: $Q\equiv
\int_{\Sigma} \ast j=f|_{\infty}$ [3]} :
\begin{equation}
Q=\rho|_{\infty}
\end{equation}
and consequently for the boundary conditions (18) the total charge
 $Q=-{\alpha^{2} }$.

\bigskip

   {\bf 8.} To calculate the ADM mass for the black hole solution (9), (10)
let us assume [3] that only the equation for $\omega$ (2) is satisfied and
consider the zweibein energy-momentum one-form $T^{a}=T^{a}_{\mu}
dz^{\mu}$ which can be determined as follows: $\delta_{e} S=\int -\ast T^{a}
\wedge \delta e^{a}$. For action (1) it takes the form:
\begin{center}
$\tilde{T}^{a} \equiv -\ast T^{a} = \alpha^{2} \nabla q^{a} + {1 \over
2}[\rho^{2} + \alpha^{2} q^{2} -
{\alpha^{4} }] \varepsilon^{a}_{\ b} e^{b}$ \\
\end{center}

Multiplying this expression on $q^{a}exp(-{ \rho \over \alpha^{2}})$
we obtain that
\begin{eqnarray}
&&T=\tilde{T}^{a}q^{a} exp(-{ \rho \over \alpha^{2}}) \nonumber \\
&&=\alpha^{2} exp(-{ \rho \over \alpha^{2}}) ( {1 \over 2} d q^{2} -
{1 \over 2\alpha^{4}} ( \rho^{2} + \alpha^{2} q^{2} - {\alpha^{4} } )
d \rho)
\end{eqnarray}
is  obviously conserved: $dT=0$.
It implies that there exist such a scalar function $m$ that
\begin{equation}
T=dm
\end{equation}
Straightforward calculations show that the mass function $m$ at point $\rho$
can be written in the following explicit form:
\begin{equation}
m={\alpha^{2} \over 2} exp(-{ \rho \over \alpha^{2}}) (q^{2} +{1 \over
\alpha^{2}}
(\rho + { \alpha^{2} })^{2})
\end{equation}
In the case when the field equations $\ast T^{a}=0$ are satisfied eq.(22)
implies that $m=constant$ and for $q^{2}(\rho)$ in the form (5) we get
that $m={\alpha^{2} \epsilon  \over 2}$.

   In order to see that defined in such way $m$ is indeed ADM mass let us
consider
the metrical energy-momentum tensor $T_{\mu \nu}={ 1 \over 2}
(T^{a}_{\mu}e^{a}_{\nu}
+T^{a}_{\nu}e^{a}_{\mu})$, $T^{a}_{\mu}=- \varepsilon_{\mu}^{\ \alpha} \tilde{
T}^{a}_{\alpha}$. Then for component $T_{00}=T^{a}_{0}e^{a}_{0}$ since
$e^{a}_{0}=q^{a}\exp({-\rho \over \alpha^{2}})$ we obtain that
\begin{eqnarray}
&&T_{00}=- \varepsilon_{0}^{\ \alpha} \tilde{T}^{a}_{\alpha} q^{a} \exp({-\rho
 \over \alpha^{2}}) \nonumber \\
&&=-\varepsilon_{0}^{\ \alpha} \partial_{\alpha} m
\end{eqnarray}
where $m$ takes the form (23). The ADM mass is ordinary determined [2,3] as
integral of $T_{00}$ over space-like hypersurface $\Sigma$ with tangent vector
$v^{\mu}=\delta^{\mu}_{1}$ and normal $n^{\mu}= \varepsilon^{\mu}_{\ \alpha}
v^{\alpha}$:
\begin{center}
$M=\int\limits_{\Sigma}^{} T_{00} n^{0} d\rho$ \\
\end{center}
One can see from (24) that it is reduced to a surface term
\begin{center}
$M=m(\rho)|_{\rho=-{\alpha^{2} \over 2}} ={\alpha^{2}\epsilon \over 2}$ \\
\end{center}

Hence only the solution of the type
{\bf A} describes the positive mass configuration (solutions of the type
{\bf B} and {\bf C} have correspondingly zero and negative mass).

\bigskip

   {\bf 9.} In conclusion we considered the two-dimensional gauge gravity
of de Sitter group (generalization to the Poincare gravity is straightforward)
and shown that the classical equations are exactly integrated in coordinate
system determined by components of 2D torsion $q^{a}, a=0,1$. The general
solution is two-dimensional asymptotically de Sitter space and for some
choice of integrating constant it turns out to be of the charged black hole
type. The square of torsion $q^{2}=q^{a}q^{b} \eta_{ab}$ is shown to be
function of curvature $\rho$ and zeros of $q^{2}(\rho)$ are points of horizons.
We calculate the conserved charge corresponding to the local Lorentz
symmetry and ADM mass which is positive for the black hole type solution.

   As the next step it would be of interest to consider the coupling of
gauge gravity with matter and analyze the Hawking radiation and back-
reaction for this type of black holes in spirit of paper [5].
This work is in progress.

   {\bf Acknowledgments.}

   I would like to express my thanks to Yu.N.Obukhov for reading the manuscript
and discussing some of the issues analyzed in this paper. I also thank to
D.Fursaev for warm hospitality at JINR in Dubna.

   {\bf Note added:}   Recently I learned that it was demonstrated in [9] that
2D $R^{2}$-gravity
with torsion is renormalizable theory. The one loop renormalizability of the
model
(11) was also proved earlier in [10].

\end{document}